%% file: ms_2c.tex
\documentclass[12pt,preprint]{emulateapj}

\newcommand{\mic}{~$\mu$m}
\newcommand{\micJy}{~$\mu$Jy}
\newcommand{\spitzer}{{\it Spitzer}}

\begin{document}

\title{Exploring the evolutionary paths of the most massive galaxies since z$\sim$2}

\author{Pablo G. P\'erez-Gonz\'alez\altaffilmark{1,2}, Ignacio Trujillo\altaffilmark{3}, Guillermo Barro\altaffilmark{1}, Jes\'us Gallego\altaffilmark{1}, Jaime Zamorano\altaffilmark{1}, Christopher J. Conselice\altaffilmark{4}}


\altaffiltext{1}{Departamento de Astrof\'{\i}sica, Facultad de CC. F\'{\i}sicas, Universidad Complutense de Madrid, E-28040 Madrid, Spain}
\altaffiltext{2}{Associate Astronomer at Steward Observatory, The University of Arizona}
\altaffiltext{3}{Instituto de Astrof\'{\i}sica de Canarias, V\'{\i}a L\'actea s/n, E-38200, La Laguna, Tenerife, Spain}
\altaffiltext{4}{University of Nottingham, School of Physics \& Astronomy, Nottingham NG7 2RD, UK}
\email{pgperez@astrax.fis.ucm.es}

\begin{abstract}

We use {\it Spitzer} MIPS data from the FIDEL Legacy Project in the
Extended Groth Strip to analyze the stellar mass assembly of massive
($\mathcal{M}$$>$10$^{11}$~$\mathcal{M}_\sun$) galaxies at z$<$2 as a
function of structural parameters. We find 24\mic\, emission for more
than 85\% of the massive galaxies morphologically classified as disks,
and for more than 57\% of the massive systems morphologically
classified as spheroids at any redshift, with about 8\% of sources
harboring a bright X-ray and/or infrared emitting AGN. More
noticeably, $\sim$60\% of all compact massive galaxies at z$=$1--2 are
detected at 24\mic, even when rest-frame optical colors reveal that
they are dead and evolving passively. For spheroid-like galaxies at a
given stellar mass, the sizes of MIPS non-detections are smaller by a
factor of $\sim$1.2 in comparison with IR-bright sources. We find that
disk-like massive galaxies present specific SFRs ranging from 0.04 to
0.2~Gyr$^{-1}$ at z$<$1 (SFRs ranging from 1 to
10~$\mathcal{M}_\sun$~yr$^{-1}$), typically a factor of 3--6 higher
than massive spheroid-like objects in the same redshift range. At
z$>$1, and more pronouncedly at z$>$1.3, the median specific SFRs of
the disks and spheroids detected by MIPS are very similar, ranging
from 0.1 to 1~Gyr$^{-1}$
(SFR$=$10--200~$\mathcal{M}_\sun$~yr$^{-1}$). We estimate that massive
spheroid-like galaxies may have doubled (at the most) their stellar
mass from star-forming events at z$<$2: less than 20\% mass increase
at 1.7$<$z$<$2.0, up to 40\% more at 1.1$<$z$<$1.7, and less than 20\%
additional increase at z$<$1. Disk-like galaxies may have tripled (at
the most) their stellar mass at z$<$2 from star formation alone: up to
$\sim$40\% mass increase at 1.7$<$z$<$2.0, and less than 180\%
additional increase below z$=$1.7 occurred at a steady rate.

\end{abstract}



\section{Introduction}

The formation and evolution of massive
($\mathcal{M}$$>$10$^{11}$~$\mathcal{M}_\sun$) galaxies is one of the
most studied topics in extragalactic astronomy during the last
decade. Early expectations from hierarchical galaxy formation models
considered that star formation began in low mass systems which built
more massive galaxies through sequential merging
\citep{1996MNRAS.283.1361B,2000MNRAS.319..168C}, in a similar process 
to the growth of structures in Cold Dark Matter simulations
\citep{2005Natur.435..629S}. In apparent contradiction with 
hierarchical assembly, the finding of a substantial population of
massive galaxies at z$>$1
\citep{1988ApJ...331L..77E,1998Natur.394..241H, 2003ApJ...587L..79F,
2004Natur.430..181G}, some of them containing old stellar populations
and evolving passively (according to their red optical colors
--\citealt{2004ApJ...617..746D}, \citealt{2005ApJ...633..748R}--, and
spectra
--\citealt{2006ApJ...649L..71K,2008ApJ...677..219K,2008A&A...482...21C}--),
seems to favor a downsizing formation scenario
\citep{1996AJ....112..839C,2004Natur.428..625H,2005ApJ...619L.135J,
2005ApJ...630...82P,2006ApJ...651..120B}. This population of z$>$1
massive galaxies accounts for a significant fraction of the
local stellar mass density ($\sim$20\% as early as z$\sim$2, and
$\sim$10\% at z$\sim$4:
\citealt{2006A&A...459..745F,2007A&A...476..137A,2008ApJ...675..234P}). 
The discovery of such a population reinforced the idea that both stars
and their host galaxies are coeval (resembling a monolithic--like
collapse), and consequently, no expectations of structure evolution in
these galaxies should be expected. For this reason, the recent
observational evidence showing that the most massive galaxies were
much more compact in the past
\citep{2005ApJ...626..680D,2006MNRAS.373L..36T,
2007MNRAS.374..614L,2008A&A...482...21C} has been surprising, and has
again opened the question of how the stellar populations of these
galaxies were assembled into their present shape.

The size evolution of the most massive objects since z$\sim$2 has been
characterized by \citet{2007MNRAS.382..109T}. These authors found
that, at a given stellar mass, disk--like objects at z$\sim$1.5 were a
factor of two smaller than their present-day counterparts. For
spheroid--like objects, the evolution has been even stronger: they
were a factor of four smaller at z$\sim$1.5 than nearby similar mass
ellipticals. In addition, the stellar mass densities of these high--z
galaxies were almost two orders of magnitude higher than objects of
the same mass today. These superdense galaxies have been found at even
higher (z$\sim$2.5) redshifts
\citep{2007ApJ...656...66Z,2007ApJ...671..285T,2008ApJ...677L...5V},  
adding more controversy to the debate about the formation and
evolution of massive galaxies.

Two processes have been proposed to allow the superdense high-z
galaxies migrate to the local stellar mass--size relation. The first
process is dissipationless (with absence of star formation) merging.
Given the high metal abundances and old ages of the stellar population
present in local massive elliptical galaxies
(e.g., \citealt{2006MNRAS.370.1106G,2006A&A...457..809S,2007ApJ...669..947J}),
these mergers should preferentially be dry
\citep{2006MNRAS.366..499D}, and occur between z$\sim$1.5 and z$=$0,
the epoch when the red sequence appears
\citep{2007ApJ...665..944L}. In this context, a particular effective
size evolutionary mechanism (r$_e$$\sim$$\mathcal{M}$$^{1.3}$) has
been provided by \citet{2006MNRAS.369.1081B} through head--on mergers
of galaxies. The second possibility is the smooth envelope accretion
scenario \citep{2007ApJ...658..710N}, where accreted stars (mainly
provided by minor mergers) form an envelope whose size increases
smoothly at decreasing redshift.

The goal of this paper is to explore the evolutionary paths followed
by the most massive galaxies and their dependence on the morphology. To
do this, we quantify the growth in stellar mass via star formation
events of massive ($\mathcal{M}$$>$10$^{11}$~$\mathcal{M}_\sun$)
galaxies as a function of size and brightness profile shape up to
z$\sim$2. We base our discussion on the characterization of the dust
infrared (IR) emission of these systems, which is linked to the amount
of recent star formation and/or the presence of obscured AGN. This
IR-based study is complementary to the more classical approach to the
characterization of the evolution of massive ellipticals based on
rest-frame optical properties.

Throughout this paper, we use a cosmology with $\mathrm
H_{0}=70$~km\,s$^{-1}$\,Mpc$^{-1}$, $\Omega_{\mathrm M}=0.3$ and
$\Lambda=0.7$. All magnitudes refer to the AB system. The results for
stellar masses and SFRs assume a \citet{2003ApJ...586L.133C} initial
mass function (IMF) with 0.1$<$$\mathcal{M}$$<$100~$\mathcal{M}_\sun$.

\section{Data description}
\label{data}

\subsection{The sample}

To analyze the star formation properties of the most massive galaxies
as a function of morphology, we use the catalog of 831 $K$-band
selected massive galaxies
($\mathcal{M}$$>$10$^{11}$~$\mathcal{M}_\sun$) in the Palomar
Observatory Wide-Field Infrared (POWIR)/DEEP-2 survey
\citep{2003SPIE.4834..161D,2006ApJ...651..120B,2008MNRAS.383.1366C} for which
\citet{2007MNRAS.382..109T} provide redshifts, stellar masses, and 
structural parameters (sizes and S\'ersic indices). These data,
jointly with the \spitzer/MIPS fluxes measured in the observations
carried our by the FIDEL Legacy Program in the Extended Groth Strip
(EGS), allow a detailed analysis of the star formation properties of
the most massive galaxies as a function of morphology up to z$\sim$2.


The sample is described in detail in \citet{2007MNRAS.381..962C} and
\citet{2007MNRAS.382..109T}. Briefly, the $K$-band survey covers
2165~arcmin$^2$ in the EGS and has a depth $K$$=$22.5~mag
(5$\sigma$). Only 710~arcmin$^2$ are covered simultaneously with
HST/ACS $v$- and $i$-band imaging from the All-Wavelength Extended
Groth Strip International Survey \citep[AEGIS,
][]{2007ApJ...660L...1D}, so reliable structural parameters can only
be measured for 831 galaxies within the entire POWIR/DEEP-2 survey in
EGS. Half of those 831 galaxies has spectroscopic redshifts based on
optical data obtained by the DEEP-2 Galaxy Redshift survey
\citep{2003SPIE.4834..161D}. \citet{2007MNRAS.381..962C} estimate 
photometric redshifts for the rest of sources with an accuracy $\Delta
z/(1+z)$$=$0.025. The 831 galaxies with
$\mathcal{M}$$>$10$^{11}$~$\mathcal{M}_\sun$ in the EGS lie in the
redshift range 0.2$<$z$\lesssim$2. Stellar masses were estimated by
\citet{2005ApJ...625..621B,2006ApJ...651..120B} using the exponential
star formation models of \citet{2003MNRAS.344.1000B} with a
\citet{2003ApJ...586L.133C} IMF and various ages, metallicities and
dust contents included. As shown by \citet{2007MNRAS.381..962C},
typical uncertainties in the stellar masses are a factor of $\sim$2
(typical of any stellar population study; see, e.g.,
\citealt{2003MNRAS.338..525P},
\citealt{2003MNRAS.341...33K},
\citealt{2006ApJ...640...92P}, \citealt{2006A&A...459..745F}, and 
\citealt{2008ApJ...675..234P}). As discussed in detail in \citet{2007MNRAS.381..962C}, 
this factor includes the effects of the photometric redshift
uncertainties, the errors inherent to solution degeneracies, and the
choices of the IMF and the stellar emission library. For example,
using \citet{2005MNRAS.362..799M} models (with an improved treatment
of the TP-AGB stellar evolution phase) would produce a 20\% (at most)
systematic decrease in the mass estimations. Using a
\citet{1955ApJ...121..161S} IMF would increase the stellar masses 
by a constant factor of 0.25dex.

\citet{2007MNRAS.382..109T} estimated (circularized) half-light radius 
($r_e$) and \citet{1968adga.book.....S} indices ($n$) for all the
galaxies in our sample. They used $i$-band HST/ACS images to fit
surface brightness profiles and divided the sample in disk-like and
spheroid-like galaxies according to the value of the S\'ersic
index. \citet[][see also
\citealt{1995MNRAS.275..874A}]{2004ApJ...604L...9R} demonstrated that
nearby galaxies with $n$$<$2.5 are mostly disks, while spheroids are
characterized by high S\'ersic indices, $n$$>$2.5. They also performed
simulations to check that the S\'ersic index obtained from HST data
can be used as a morphology indicator at
z$>$0. \citet{2007MNRAS.382..109T} extended these simulations to prove
that the structural parameters are robust against the effects
introduced by $K$-corrections, AGN contamination, and surface
brightness dimming.

For our sample, visual inspection of the ACS $i$-band images by one of
the co-authors (I.T.) was used to classify the sample in 4 types:
ellipticals/lenticulars, spirals, irregulars, and mergers. Comparing
this visual classification with the one based on S\'ersic indices, we
find that the visually confirmed spheroids present
$<$$n$$>$$=$4.8$\pm$1.5, and the rest of sources have
$<$$n$$>$$=$1.7$\pm$1.7. There is a 6\% contamination of visually
identified spheroids in the $n$$<$2.5 sample, and a 20\% contamination
of visually identified disks in the $n$$>$2.5 sample (comparable to
the 5\% and 19\% contaminations in \citealt{2004ApJ...604L...9R}). The
fraction of interloper disks decreases to 7\% at $n$$>$4 and 4\% at
$n$$>$5. A 4\% contamination is typical of other works based on visual
or quantitative morphological classifications such as
\citet{2007MNRAS.381..962C} or \citet{2005ApJ...625..621B}. 
The visual test shows that the galaxies with $n$$>$4 form a robust
(almost uncontaminated) sample of spheroid-like sources, and $n$$<$2.5
galaxies are mostly disks. Galaxies with 2.5$<$n$<$4 are most probably
spheroid-like galaxies with some contamination of S0 galaxies and
early-type spirals.

\subsection{UV-to-MIR photometric properties of the sample}
\label{merged}

\begin{figure*}
\vspace{-2cm}
\begin{center}
\begin{minipage}{5cm}
\includegraphics[angle=0,width=4.cm]{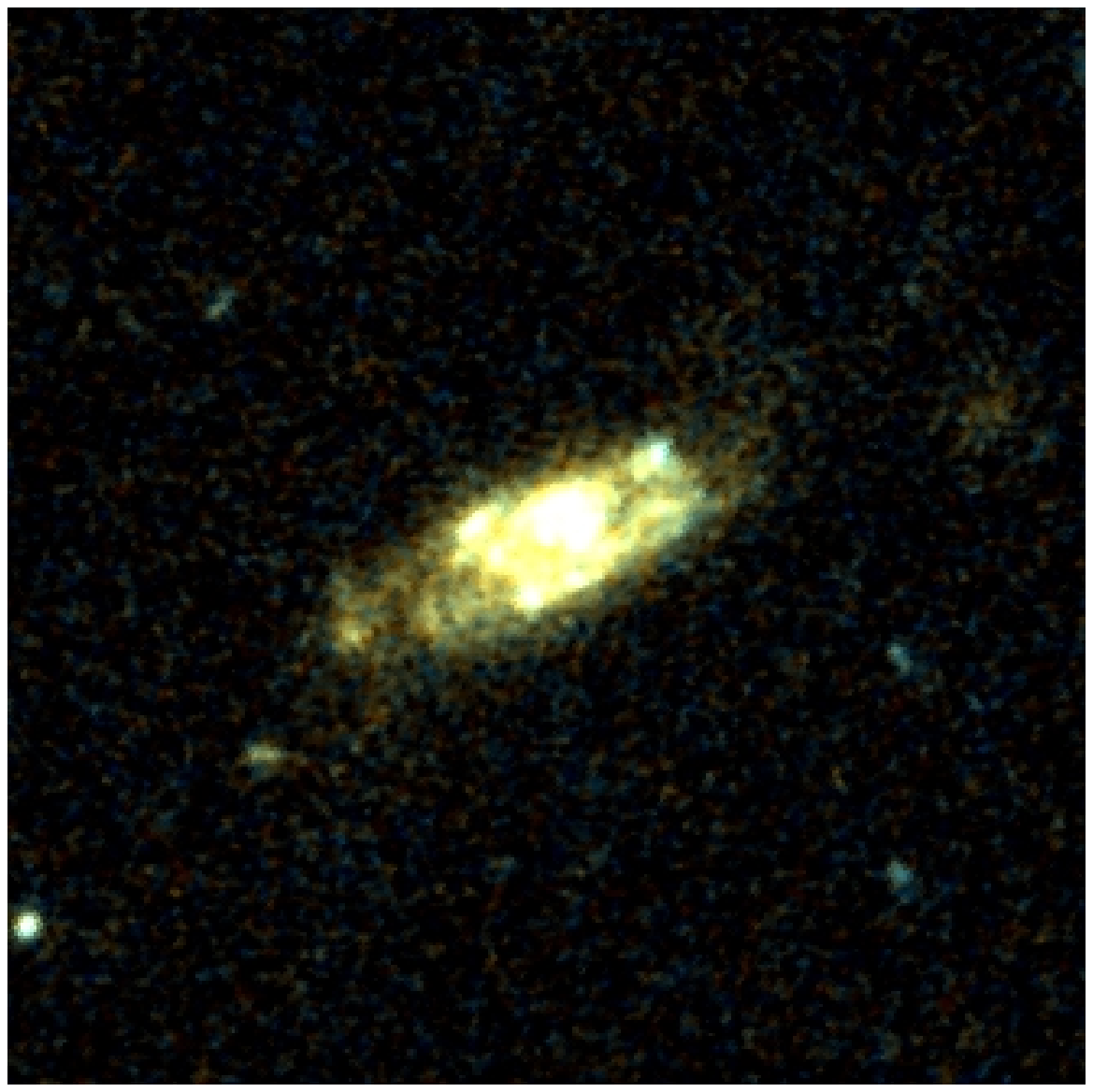}
\end{minipage}
\begin{minipage}{5cm}
\includegraphics[angle=-90,width=4.cm]{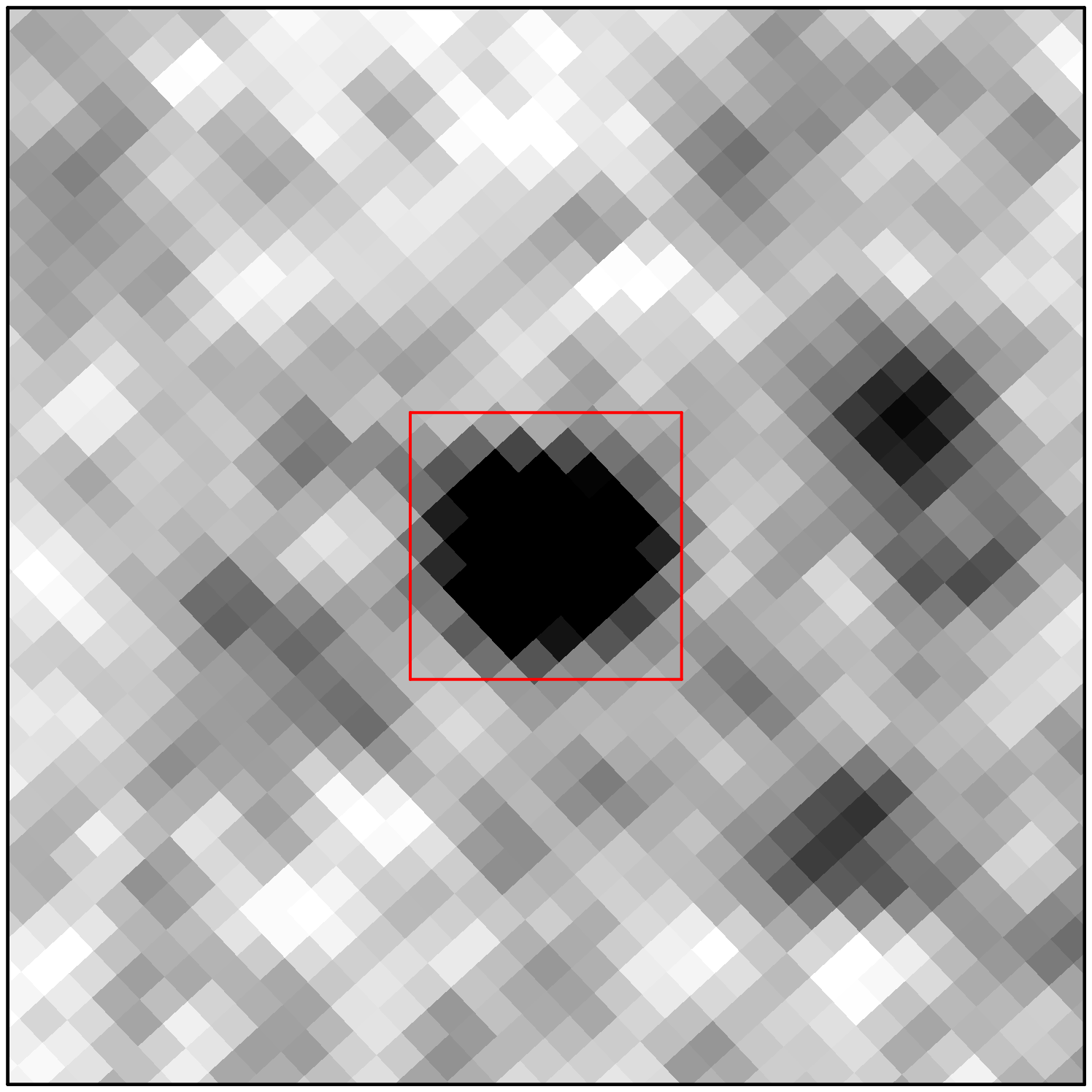}
\end{minipage}
\begin{minipage}{6.cm}
\includegraphics[angle=-90,width=5.cm]{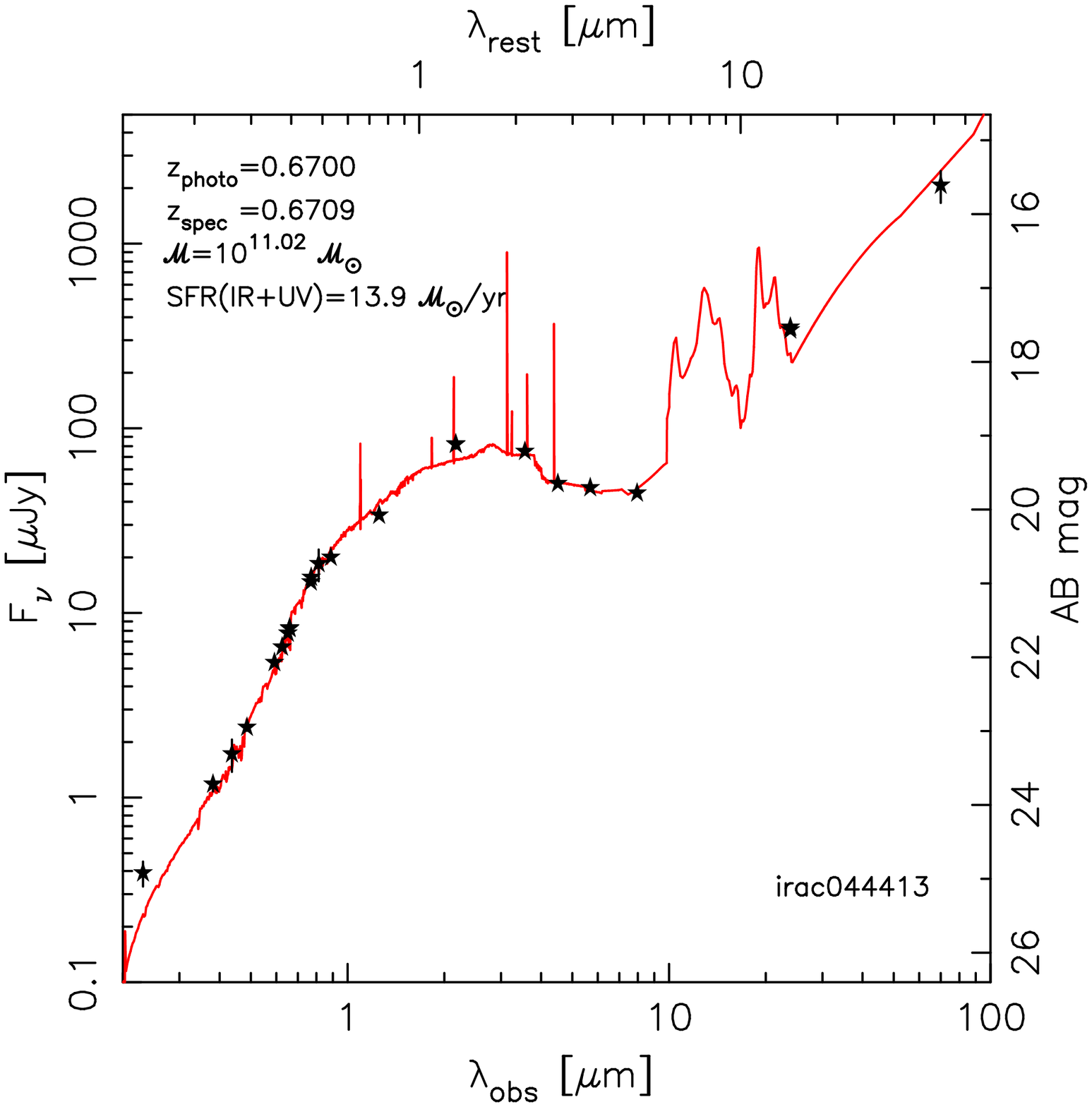}
\end{minipage}

\begin{minipage}{5cm}
\includegraphics[angle=0,width=4.cm]{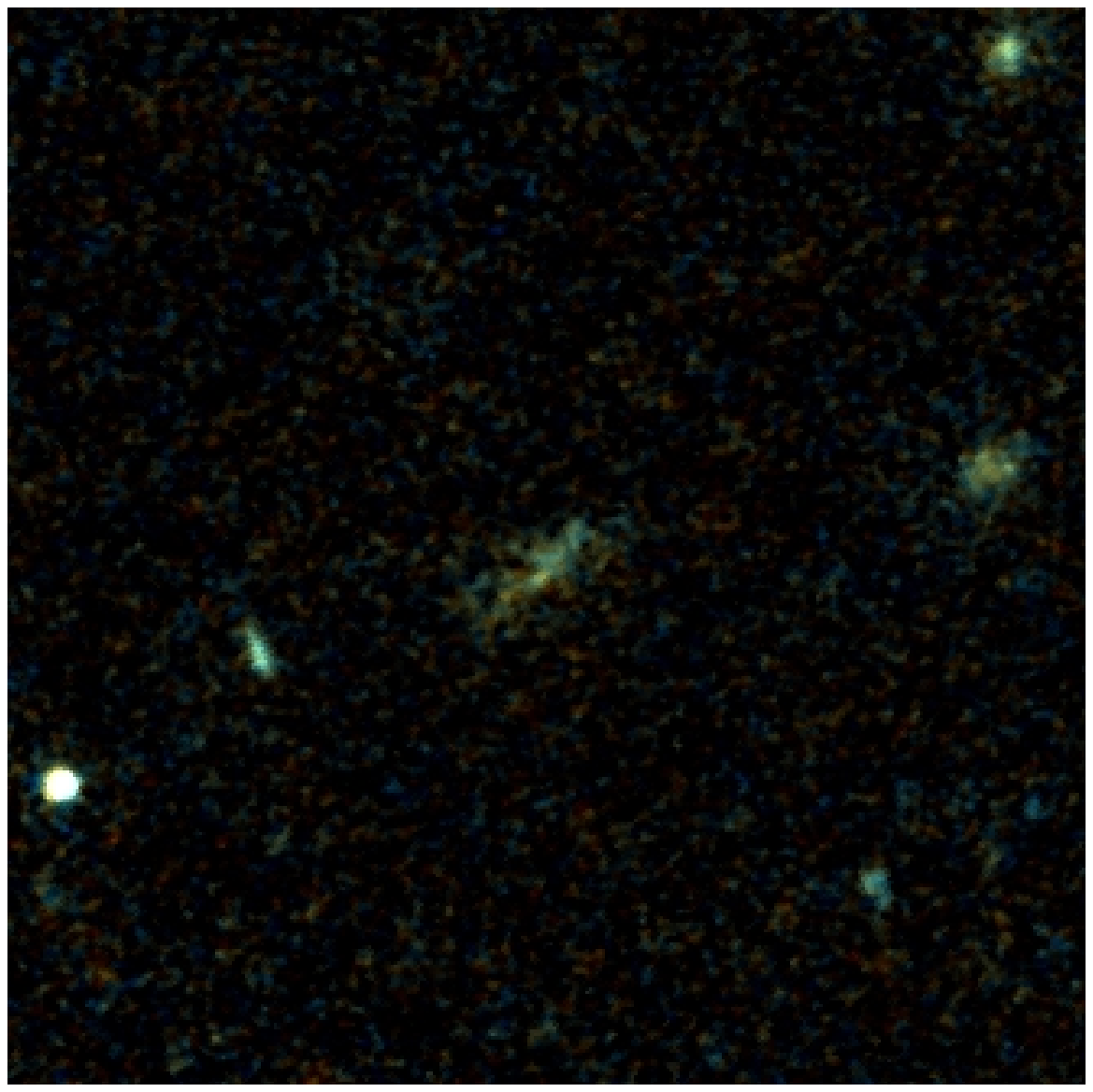}
\end{minipage}
\begin{minipage}{5cm}
\includegraphics[angle=-90,width=4.cm]{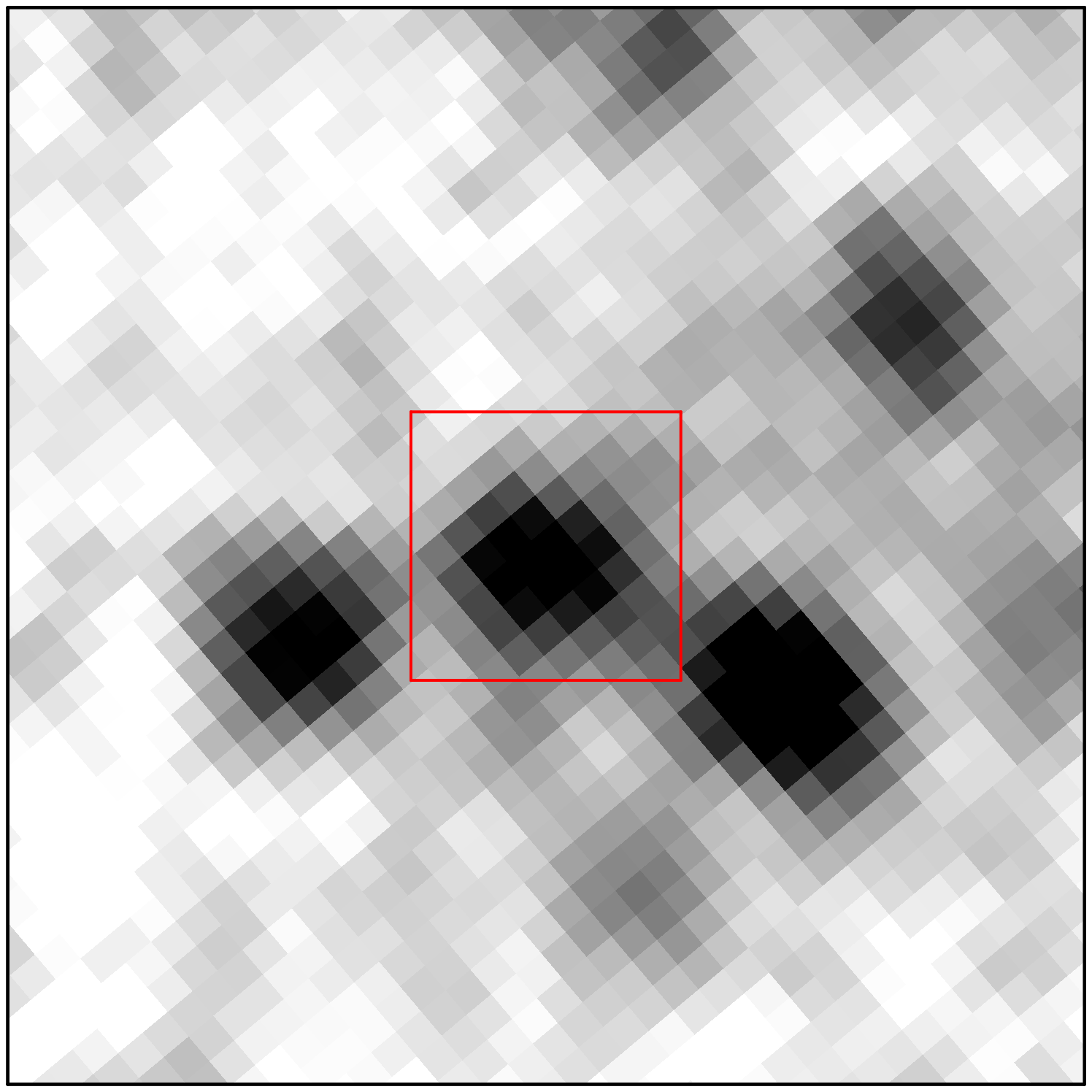}
\end{minipage}
\begin{minipage}{6.cm}
\includegraphics[angle=-90,width=5.cm]{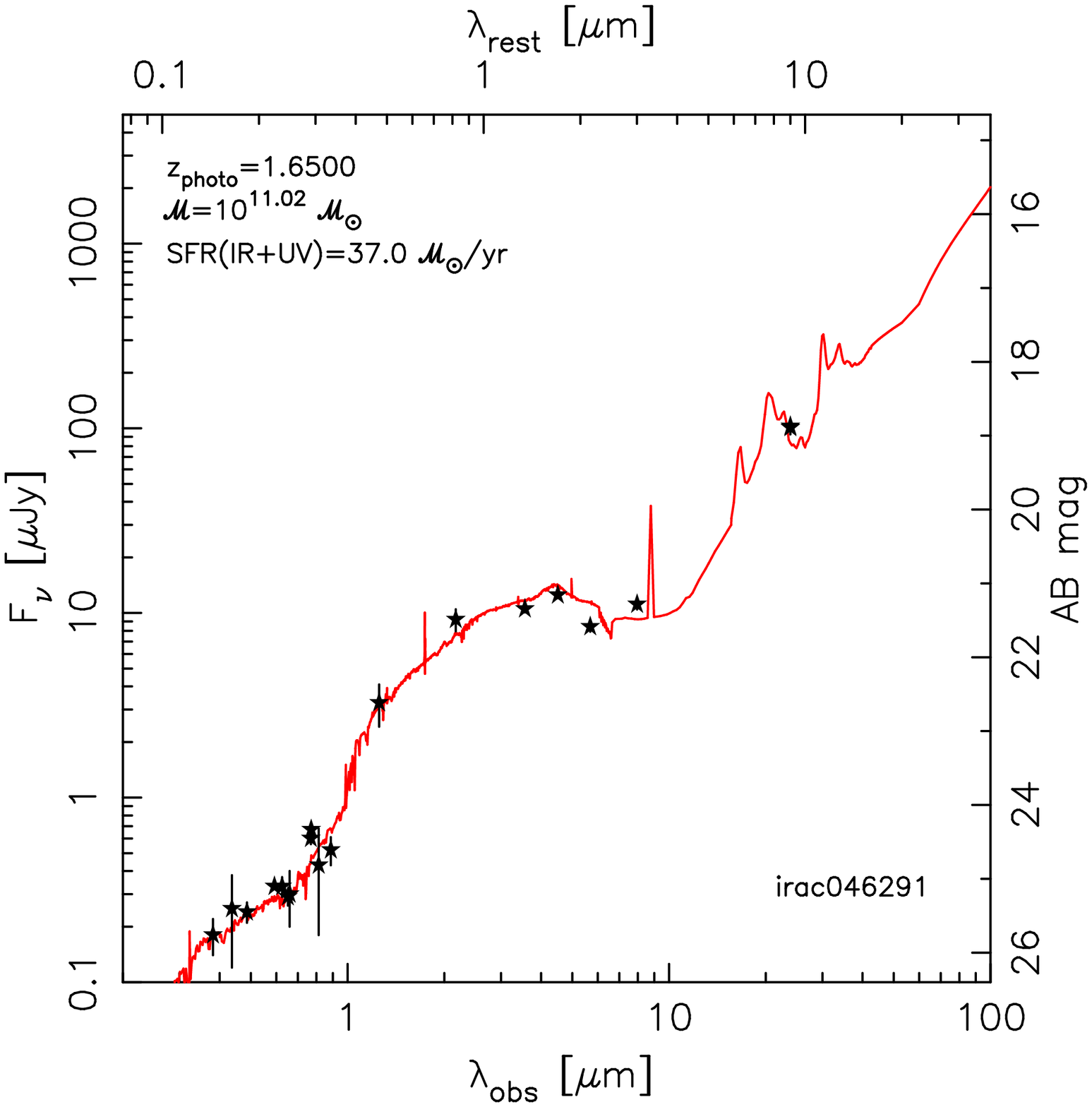}
\end{minipage}

\begin{minipage}{5cm}
\includegraphics[angle=0,width=4.cm]{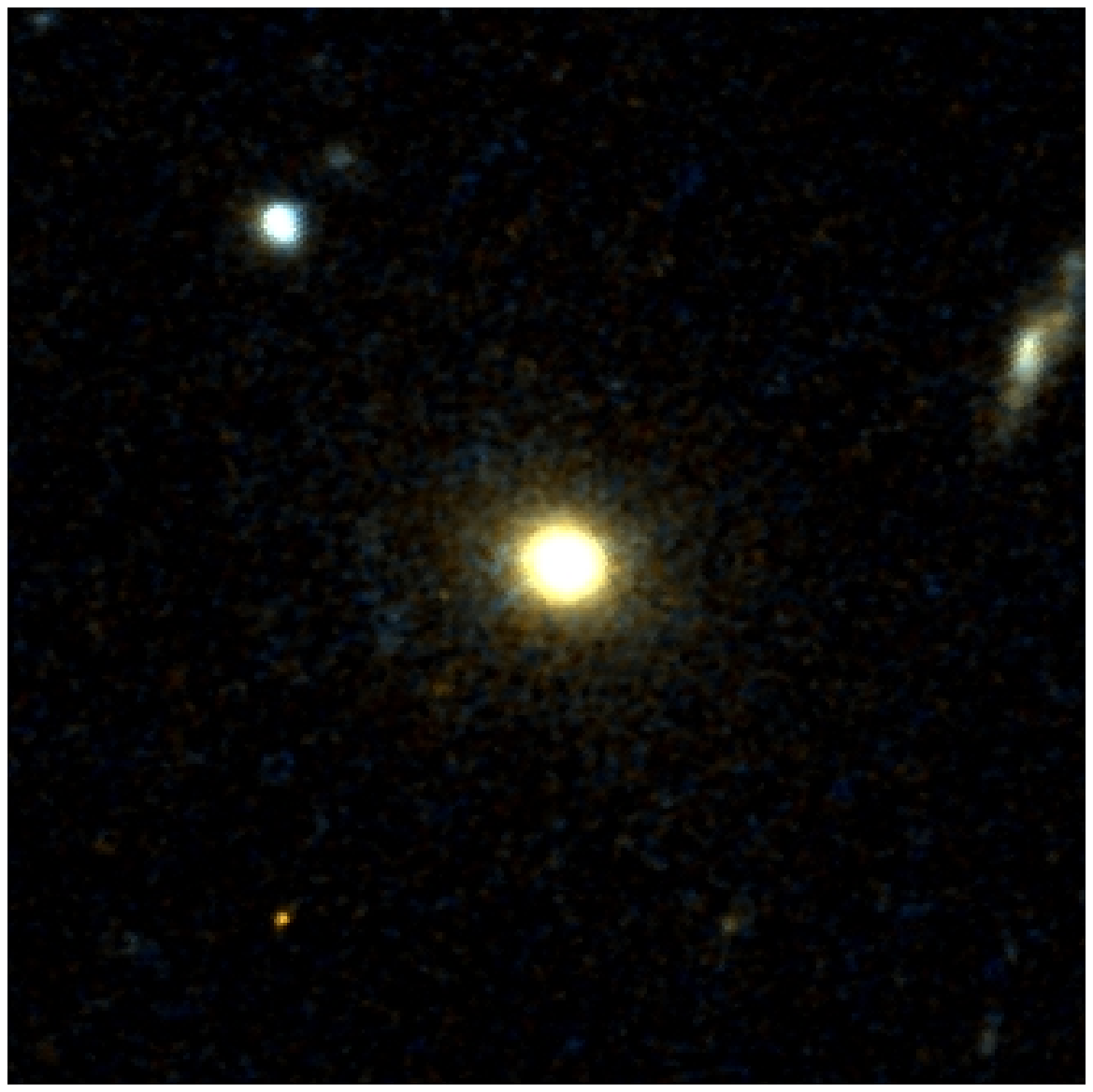}
\end{minipage}
\begin{minipage}{5cm}
\includegraphics[angle=-90,width=4.cm]{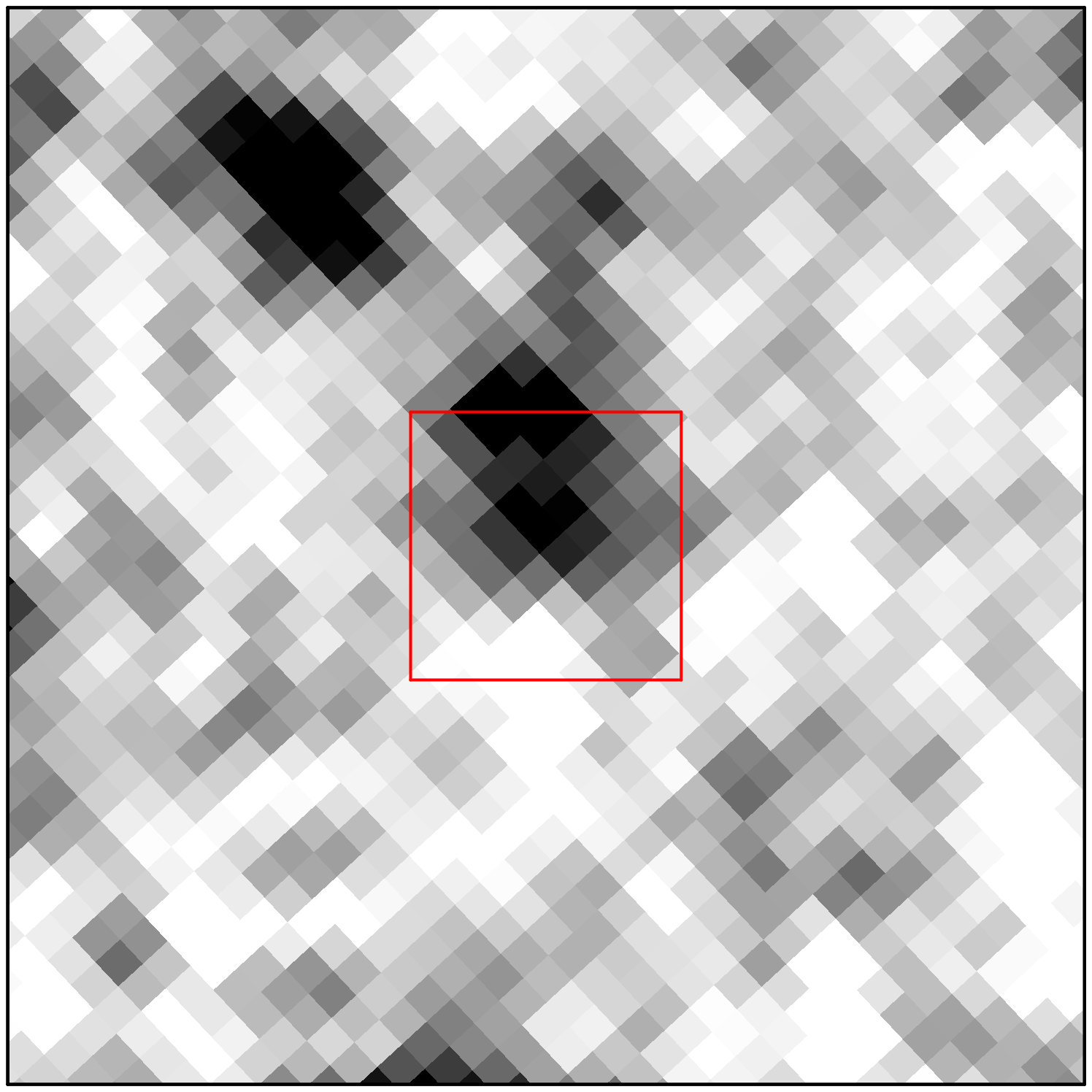}
\end{minipage}
\begin{minipage}{6.cm}
\includegraphics[angle=-90,width=5.cm]{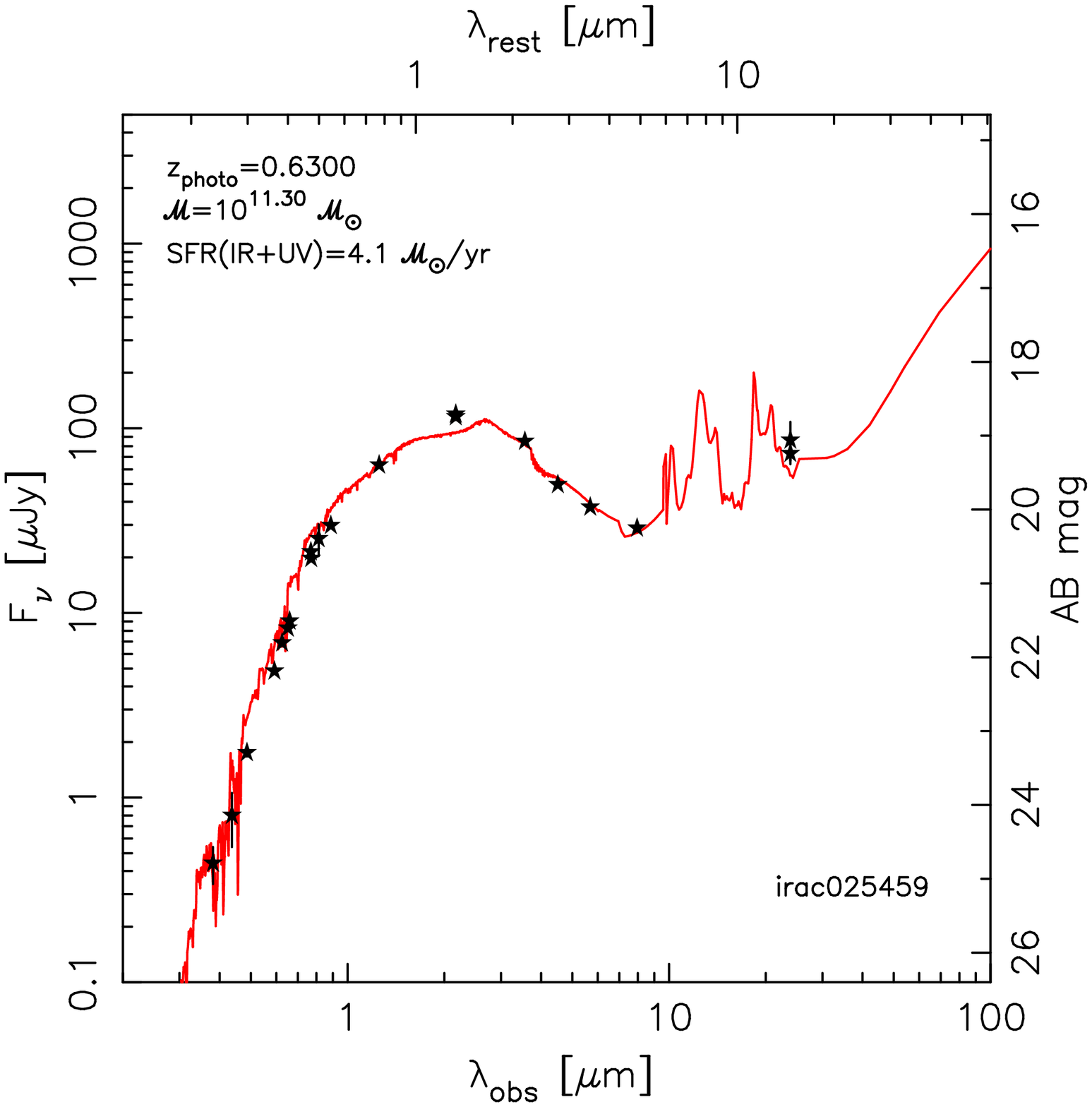}
\end{minipage}


\begin{minipage}{5cm}
\includegraphics[angle=0,width=4.cm]{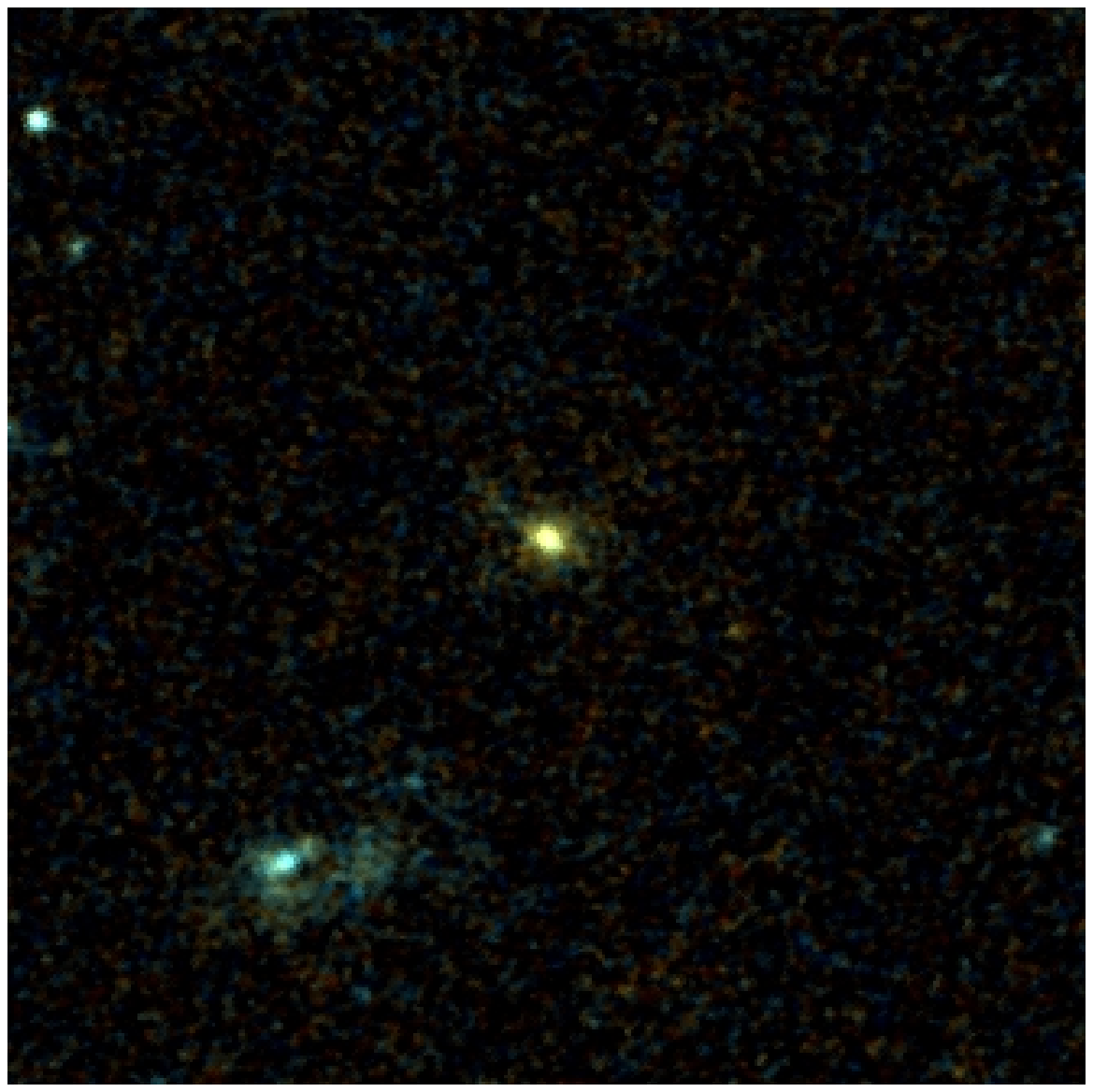}
\end{minipage}
\begin{minipage}{5cm}
\includegraphics[angle=-90,width=4.cm]{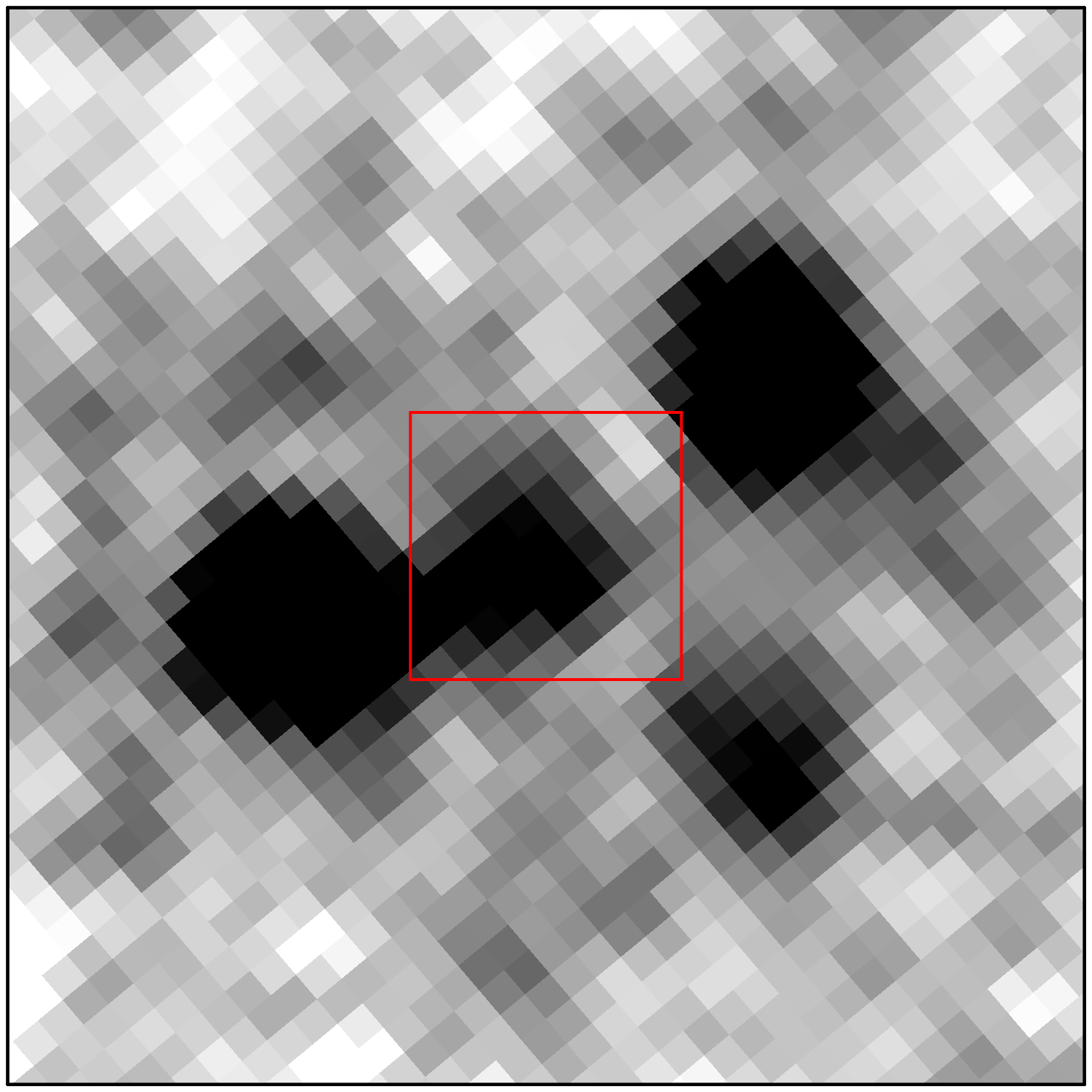}
\end{minipage}
\begin{minipage}{6.cm}
\includegraphics[angle=-90,width=5.cm]{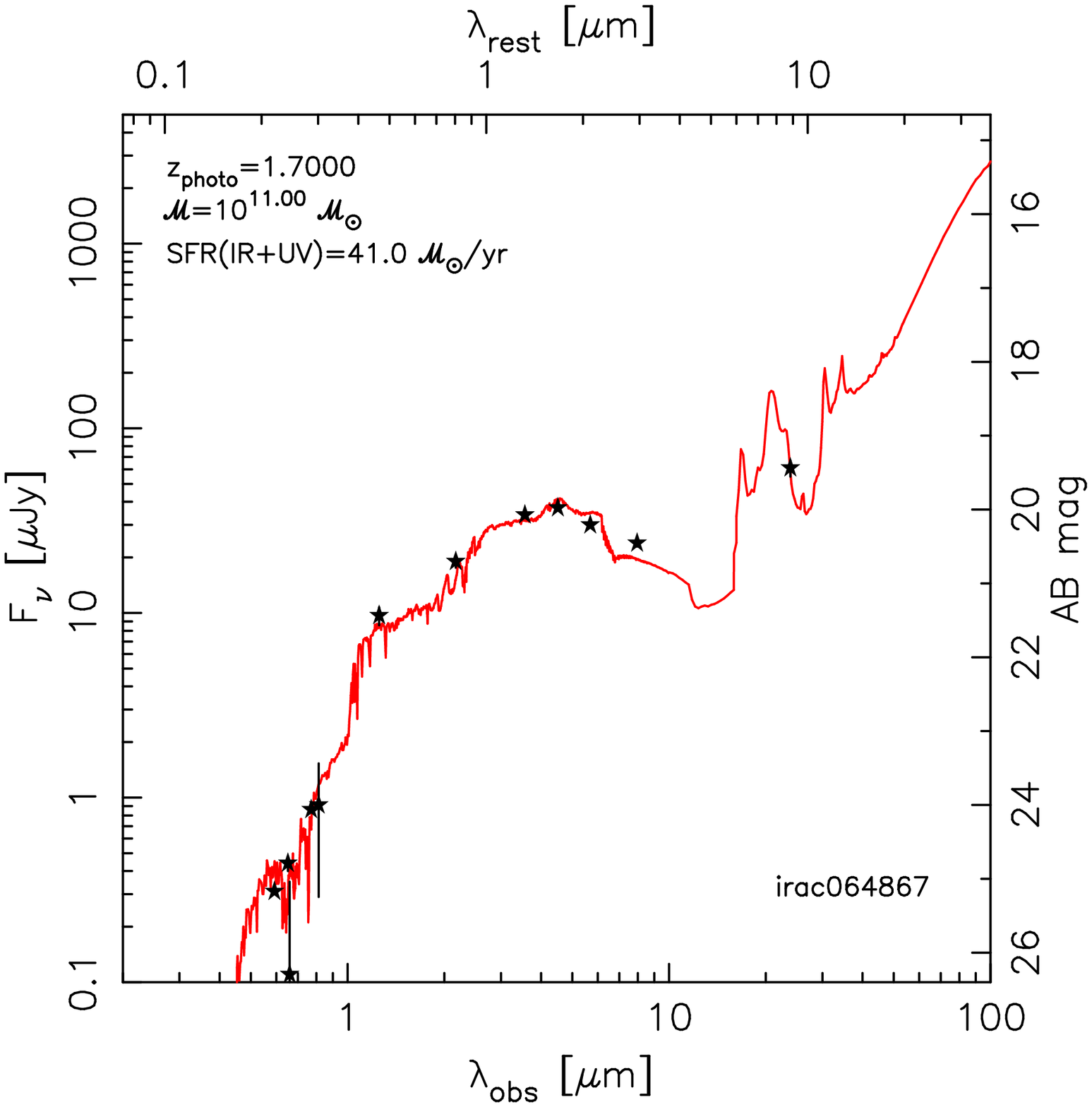}
\end{minipage}

\figcaption{\label{postage_stamps} Postage stamps and SEDs for four 
typical MIPS-detected galaxies in our sample of massive galaxies. Left
panels show 10$\arcsec$$\times$10$\arcsec$ RGB composite images built
from HST/ACS $v$ and $i$ frames. In the middle column, MIPS 24\mic\,
images of size 40$\arcsec$$\times$40$\arcsec$ are depicted, with the
red square showing the area covered by the ACS postage stamp. In all
images, North is up and East is left. The right columns show the SEDs
of each galaxy, fitted to stellar population and dust emission models
which are used to estimate photometric redshifts, stellar masses and
SFRs (these parameters are given in each SED plot). The two upper rows
show examples of disk-like galaxies: EGS142126.97$+$531137.4, a galaxy
at z$=$0.67; and EGS142013.18$+$525925.0, lying at z$=$1.65. The two
lower rows show examples of spheroid-like galaxies:
EGS142021.47$+$525543.4, a galaxy at z$=$0.63, and
EGS142125.76$+$531622.8, placed at z$=$1.70.}
\end{center}
\end{figure*}

The positions of the 831 massive galaxies in the EGS were
cross-correlated (using a 1$\arcsec$ search radius) with our own
reduction and catalogs of the {\it Spitzer} IRAC survey of the
EGS. Using the same simulation method described in
\citet{2008ApJ...675..234P}, we found that this catalog is 75\%
complete at $\sim$1.5\micJy\, ([3.6]$=$23.5~mag), which corresponds to
8$\sigma$ detections. Our IRAC photometry is consistent with that
published by \citet{2008arXiv0803.0748B} for the same dataset (but
their own reduction and cataloging) within 0.1~mag (typical absolute
uncertainty of IRAC fluxes) for 75\% of the sample, and within
1-$\sigma$ error for virtually all sources.

We found IRAC counterparts down to [3.6]$=$23~mag for all of the 831
galaxies in Trujillo's sample. For 151 sources (18\% of the total),
the IRAC sources were blended with nearby objects, but still resolved
(the separation was larger than 1$\arcsec$). As described in detail in
Appendix A of \citet{2008ApJ...675..234P}, for these sources we
obtained multi-wavelength photometry (including \spitzer\, fluxes)
using a deblending algorithm based on the deconvolution of the IRAC
and MIPS images. The method takes the known positions of the blended
sources obtained from optical/NIR ground-based images and the PSFs for
the different images and obtain separated fluxes for the blended
sources \citep[see also ][]{2006A&A...449..951G}. The method relies on
the moderate resolution of the IRAC images ($\sim$2$\arcsec$ FWHM, not
that different from an optical ground-based image, but very stable),
which allows the deblending of sources separated by more than
$\sim$1$\arcsec$ (half of the FWHM). For the MIPS 24\mic\, images, the
resolution is worse ($\sim$6$\arcsec$ FWHM) but the IRAC data can be
used to assign most probable counterparts and help with the
deblending. In any case, the main results in this paper (the MIPS
detection fraction and the statistics of specific SFRs) remain
virtually unchanged (less than 5\% random changes at all redshifts)
when we remove the 151 sources with blending problems.


We measured consistent aperture photometry in several UV, optical,
NIR, and MIR bands with the method described in
\citet{2008ApJ...675..234P}. The multi-wavelength dataset is outlined briefly in
\citet{2008ApJ...677..169V} and will be characterized in detail in
Barro et al. (2008, in preparation). More noticeably, our merged
photometric catalog includes MIPS fluxes at 24\mic\, obtained from
aperture photometry in the GTO and FIDEL survey (DR2) data in the
AEGIS/EGS field
\citep{2007ApJ...660L...1D,2007ApJ...660L..73S}. Following the same
procedure described in
\citet{2005ApJ...630...82P,2008ApJ...675..234P}, we used the {\sc DAOPHOT} 
software package in IRAF\footnote{IRAF is distributed by the National
Optical Astronomy Observatory, which is operated by the Association of
Universities for Research in Astronomy (AURA), Inc., under cooperative
agreement with the National Science Foundation.} to detect sources
(using a 3$\sigma$ detection cut above the local sky noise) in the
MIPS images and measure aperture photometry with a PSF fitting method.

Based on simulations consisting on the creation and recovery of
artificial sources in these images, we found that our 24\mic\, catalog
of the EGS is 75\% complete at 35\micJy\, (consistent with
\citealt{2004ApJS..154...70P} and \citealt{2006ApJ...640..603T}). 
$F(24)$$=$35\micJy\, corresponds to $\sim$6$\sigma$ detections for the
average sky noise in the FIDEL DR2 images. Our 24\mic\, catalogs are
cut to 3$\sigma$ detections, which translate to the range
14--17\micJy, depending on the location on the sky due to small
differences in exposure time and the effect of confusion (both
presenting small spatial variations throughout the image). As done in
\citet{2005ApJ...630...82P}, we tested the reliability of the MIPS 
detections by analyzing the probability of having a counterpart within
the search radius (1$\arcsec$) in other optical/NIR (ground-based and
IRAC) bands for a random position on the sky of the 24~\mic\,
image. Having a counterpart in 3 different bands within the EGS
dataset has a negligible probability (1.4\%), so we conclude that
(virtually) all the MIPS detections within our sample are not
spurious. Figure~\ref{postage_stamps} shows postage stamps and
spectral energy distributions (SEDs) of two typical examples of
disk-like galaxies and two spheroid-like galaxies detected by MIPS at
24\mic\, (also one 70\mic\, detection included).


Using the measured MIPS 24\mic\, fluxes, the estimated 280~nm
synthetic fluxes inter/extrapolated in the spectral energy
distribution fits, and the spectroscopic and photometric
redshifts published by \citet{2007MNRAS.382..109T}, we obtained total
(unobscured plus obscured) SFRs for each galaxy in the same way
explained in
\citet[][see also \citealt{2005ApJ...625...23B}]{2008ApJ...675..234P}. 
Briefly, the MIR fluxes at rest-frame wavelengths longer than 5\mic\,
are fitted to dust emission models (from several libraries) and the
IR-based SFRs are obtained from integrated total IR (using the
calibration in \citealt{1998ARA&A..36..189K}) and rest-frame 24\mic\,
(see \citealt{2006ApJ...650..835A}) luminosities from the fits
(averaged through all template libraries). The IR-based (obscured) SFR
is then added to the UV-based (unobscured) SFR to obtain the total
SFR. As discussed in that paper, total SFR estimates should be good
within a factor of 2. The SFRs discussed in the following Sections
were estimated assuming a
\citet{2003ApJ...586L.133C} IMF, obtained by dividing the results 
obtained with the calibrations in \citet[][valid for a
\citealt{1955ApJ...121..161S} IMF]{1998ARA&A..36..189K} by a 1.8
factor.

\section{Results and discussion}
\label{results}

\subsection{IR emission of the most massive galaxies at z$\lesssim$2}

\begin{figure*}
\begin{center}
\includegraphics[angle=0,width=17cm]{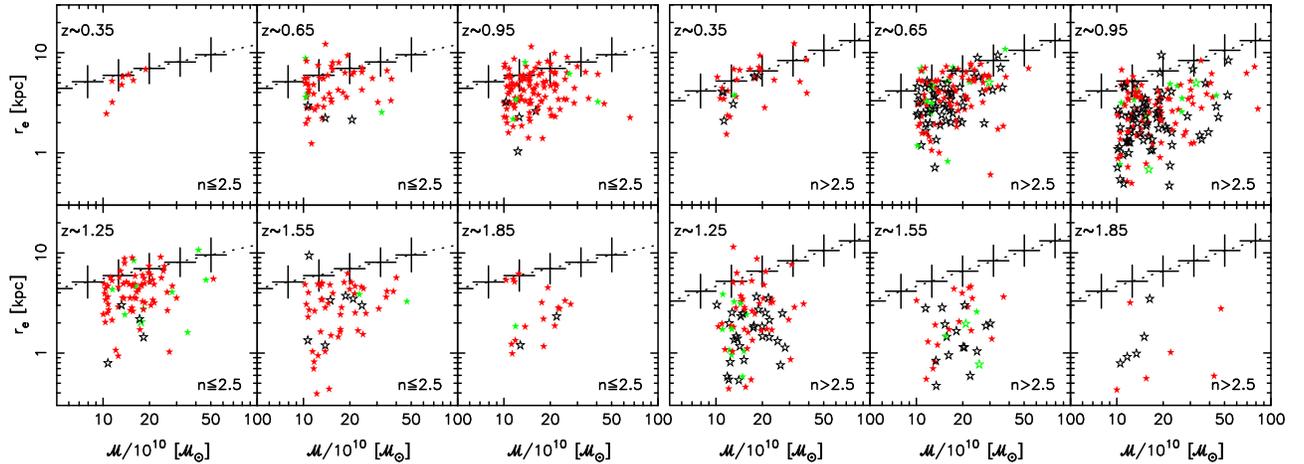}
\figcaption{\label{mass_re}Stellar mass--size distribution  for different 
redshift bins of our massive galaxies separated in disk-like (left
panels) and spheroid-like (right panels) types. Galaxies detected by
MIPS at 24\mic\, are plotted with filled stars, while open stars show
MIPS non-detections. Red symbols are galaxies whose MIPS emission is
identified with obscured star formation, and green symbols depict
galaxies which harbor an X-ray or/and IR-emitting AGN.}
\end{center}
\end{figure*}

Figure~\ref{mass_re} shows the location of the MIPS detections in a
stellar mass--size diagram for massive galaxies as a function of the
concentration index. Out of the 831 galaxies in Trujillo's sample of
massive galaxies, 485 sources (58\%) were classified as spheroid-like
based on their high ($n$$>$2.5) \citet{1968adga.book.....S} indices,
and 346 (42\%) as disk-like sources ($n$$\leq$2.5). Among the
disk-like sources, 322 (93\% of all disks) are detected by MIPS at
24\mic\, with a minimum flux $F(24)$$=$15\micJy, and 137 galaxies are
detected at 70\mic\, down to $F(70)$$=$0.5~mJy. Among the spheroids,
297 galaxies (61\% of the total) are detected at 24\mic\, down to
$F(24)$$=$14\micJy, and 84 sources are detected at 70\mic\, down to
$F(70)$$=$0.7~mJy. If we consider only the MIPS 5$\sigma$ detections
(i.e., more statistically reliable sources, although our simulations
reveal that all our 3$\sigma$ 24\mic\, sources are reliable to the
99\% level; see Section~\ref{merged}), the MIPS detection fractions
decrease to 92\% for disky systems and 52\% for spheroid-like
sources. 

\input{tab1}

\begin{figure*}
\begin{center}
\includegraphics[angle=-90,width=17.cm]{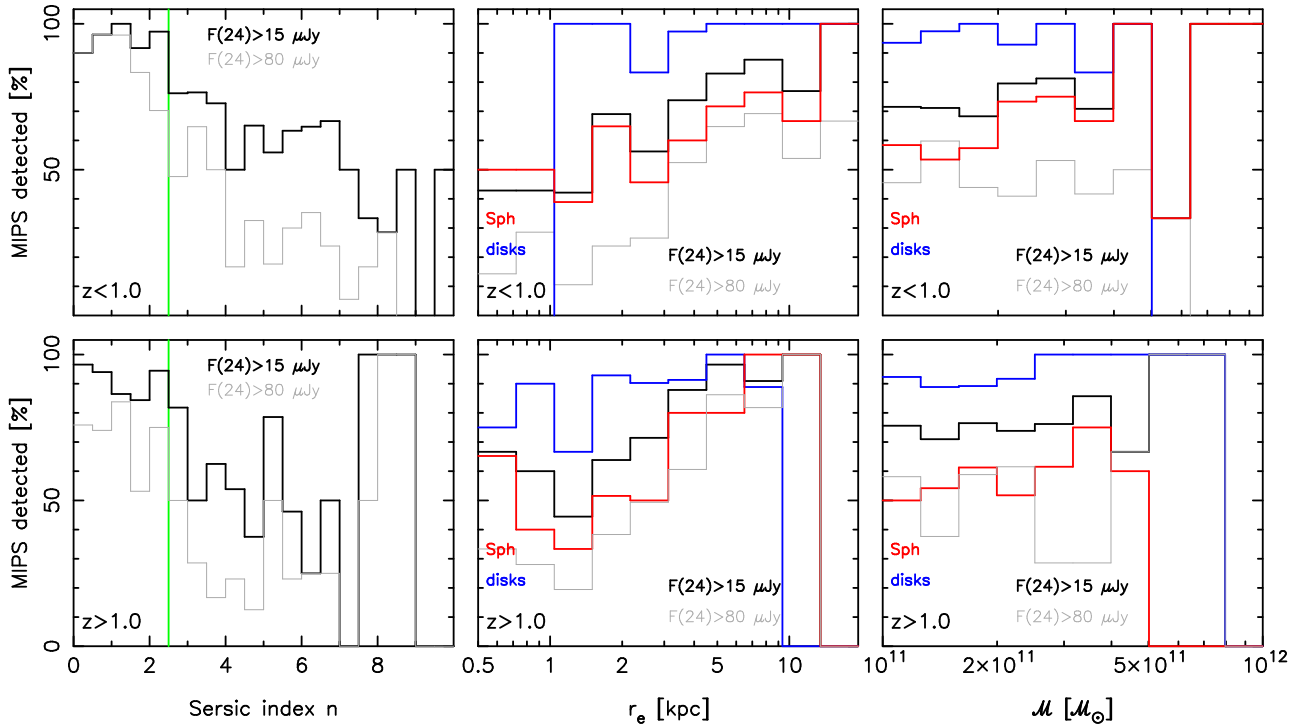}
\figcaption{\label{re_n_mips} MIPS 24\mic\, detection fractions as a 
function of S\'ersic index (left panels), (circularized) half-light
radius (middle panels), and stellar mass (right panels). The whole
sample has been divided into two redshift intervals (z$<$1 on the
top panels, and z$>$1 on the bottom panels). Galaxies identified as
IR- or X-ray-bright AGN are excluded from the distributions. In all
plots, wide black lines show the results for all the MIPS detections
(with a minimum measured value of $F(24)$$=$15\micJy) and gray narrow
lines show the measured fractions for a flux cut
$F(24)$$>$80\micJy. On the left panels, the green vertical line shows
the adopted separation between spheroid-like and disk-like
galaxies. On the middle and right panels, red lines show the MIPS
detection fractions for spheroid-like sources, and blue lines for
disk-like sources.}
\end{center}
\end{figure*}

Table~\ref{table1} shows the total number of sources and MIPS
detection fractions (for fluxes above the 3$\sigma$ level) as a
function of redshift and morphology. We consider the results based on
the morphological classification using the S\'ersic indices and the
direct visual inspection of the images. The MIPS detection fractions
for $n$$\leq$2.5 galaxies and visually identified disks are almost
identical. Visually confirmed spheroids present less MIPS detections
than the $n$$>$2.5 sources, although the difference is small ($<$10\%)
and consistent with the 20\% contamination of visually identified
disks in the $n$$>$2.5 sample, most of them having 2.5$<$$n$$<$4.0
(see Figure~\ref{re_n_mips}).

Figure~\ref{mass_re} shows that there is basically no difference
between the loci occupied by MIPS detected and undetected galaxies in
the stellar mass--size plane. However, for spheroid-like objects at a
given stellar mass, MIPS non-detections are smaller than IR-bright
sources by a factor of $\sim$1.2 (see also
\citealt{2007ApJ...656...66Z} results at higher z). This suggests that 
early (i.e., z$>$2) massive star formation events left even more
compact remnants than starbursts taking place at z$<$2, maybe
reflecting the higher density conditions of the primitive Universe.


Figure~\ref{re_n_mips} shows the MIPS 24\mic\, detection fractions as
a function of structural parameters. This Figure confirms the bias of
the $n$$\leq$2.5 sample towards galaxies with on-going (possibly
extended through the disk) star formation or harboring an IR-emitting
AGN. Figure~\ref{re_n_mips} demonstrates that virtually all (80-90\%)
the $n$$\leq$2.5 galaxies are detected by MIPS at all redshifts and
(almost) independently of the size of the galaxy. In contrast, the
spheroid sample is biased towards more quiescent systems. There is
still a non-negligible fraction (6\%) of galaxies classified as disky
which fall below the MIPS detection limit or do not present any IR
emission, i.e, they have low-level star formation, no dust, or are
completely quiescent. According to Figure~\ref{re_n_mips}, most of
them lie at z$>$1 (59\% of all disk-like non-detections) and tend to
have comparatively smaller sizes: the MIPS detection fraction
decreases from 90\% for disk-like galaxies with $r_e$$\gtrsim$4~kpc to
70-80\% for $r_e$$\lesssim$1.5~kpc systems. All these sources present
very red SEDs (see Figure~\ref{seds}).


On the contrary, the spheroid-like sample includes at least a 63\% of
''active'' galaxies. Moreover, some more MIPS undetected spheroids may
have some star formation activity or harbor an AGN, since some of the
SEDs in the upper-left panel of Figure~\ref{seds} present a
significant emission in the UV, probably arising from young
stars. Most of these UV-bright galaxies lie at z$>$1, and the MIPS
24\mic\, flux upper limits\footnote{The flux upper limits at 70~$\mu$m
have been omitted from Figure~\ref{seds} for clarity, given that very
few sources are detected at this wavelength.}  are consistent with the
MIR emission from a typical Sc galaxy. Figure~\ref{re_n_mips} shows
that MIPS detections are more common among the largest spheroid-like
galaxies, especially at z$>$1. Interestingly, at z$>$1 the MIPS
detection fraction stays roughly constant at $r_e$$\lesssim$3~kpc.  It
may even increase (up to 70\%) for very compact ($r_e$$\lesssim$1~kpc)
spheroid-like galaxies, although there are 3 caveats to this result:
1) the number of sources with $r_e$$\lesssim$1~kpc is small
($\sim$30), so the uncertainties in these bins are of the order of
20-30\%; 2) for z$>$1 and $r_e$$\lesssim$1~kpc we are reaching the
resolution limit in the HST/ACS images, and consequently the size
measurements count with large uncertainties; and 3) these high-z
compact galaxies may be dominated by an obscured AGN (since the
galaxies are detected by MIPS and show no X-ray emission; see also
\citealt{2007MNRAS.382..109T}), which may bias the size measurements.

Our results may be compared with those published by
\citet{2007MNRAS.376..416R}. They found that 20\% of the most securely 
identified spheroids\footnote{Note that \citet{2007MNRAS.376..416R}
cut their sample to clear E and E/S0 galaxies as classified visually
by
\citet{2005ApJ...625..621B}, but do not include Bundy's S0 type in
their analysis.} at 0.3$<$z$<$1.0 are detected during phases of
prominent activity. The evidence is the detection at 24\mic\, above
80\micJy\, or at radio wavelengths above 40\micJy. Our MIPS data are
deeper than theirs, but if we cut our catalogs to the same flux limit
(see Figure~\ref{re_n_mips}), we obtain an average $\sim$20\%
detection fraction at z$<$1 for the most concentrated objects with
$n$$>$4, probably well correlated with their sample of bona fide
spheroids, although our selection only includes the most massive
galaxies ($\mathcal{M}$$>$10$^{11}$~$\mathcal{M}_\sun$), and
Rodighiero's selection  is not based on stellar mass.

If we consider the high-z (z$>$1) galaxies in our sample, our results
are also consistent with those found in \citet{2006ApJ...640...92P}
for Distant Red Galaxies (DRGs). These galaxies have a typical stellar
mass $\mathcal{M}$$\sim$10$^{11}$~$\mathcal{M}_\sun$ and a mean
redshift z$\sim$2
\citep[see also][]{2007A&A...465..393G,2008ApJ...675..234P}. Given that 
our sample has a stellar mass cut
$\mathcal{M}$$>$10$^{11}$~$\mathcal{M}_\sun$ and a redshift cut at
z$=$2, the low-redshift, high-mass tail of the general DRG population
must be included in our selection. Indeed, we have 27 DRGs in our
sample, with an average redshift
$<$z$>$$=$1.50$\pm$0.22. \citet{2006ApJ...640...92P} found that
roughly 50\% of DRGs are detected by MIPS at 24\mic\, down to
80\micJy. We find a $\sim$60\% detection fraction for massive galaxies
in the low redshift tail of DRGs (1$<$z$<$2) with deeper data (75\%
completeness level at 35\micJy). Among the 27 DRGs in our sample, 25
(93\%) are detected by MIPS. This detection fraction is higher than
the average for 1$<$z$<$2 galaxies, but still consistent with the
results in
\citet{2006ApJ...640...92P}, who argue that z$\lesssim$2 DRGs are mostly 
heavily extincted starbursts part of the class of dusty EROs at z$>1$,
and find a $\sim$75\% MIPS detection fraction for their
$\mathcal{M}$$>$10$^{11}$~$\mathcal{M}_\sun$ DRGs at z$<$2. Indeed, 18
DRGs in our sample are EROs (16 MIPS detections), and we count with a
total of 305 EROs in our entire sample.

\subsection{Spectral energy distributions}


\begin{figure*}
\begin{center}
\includegraphics[angle=-90,width=15.cm]{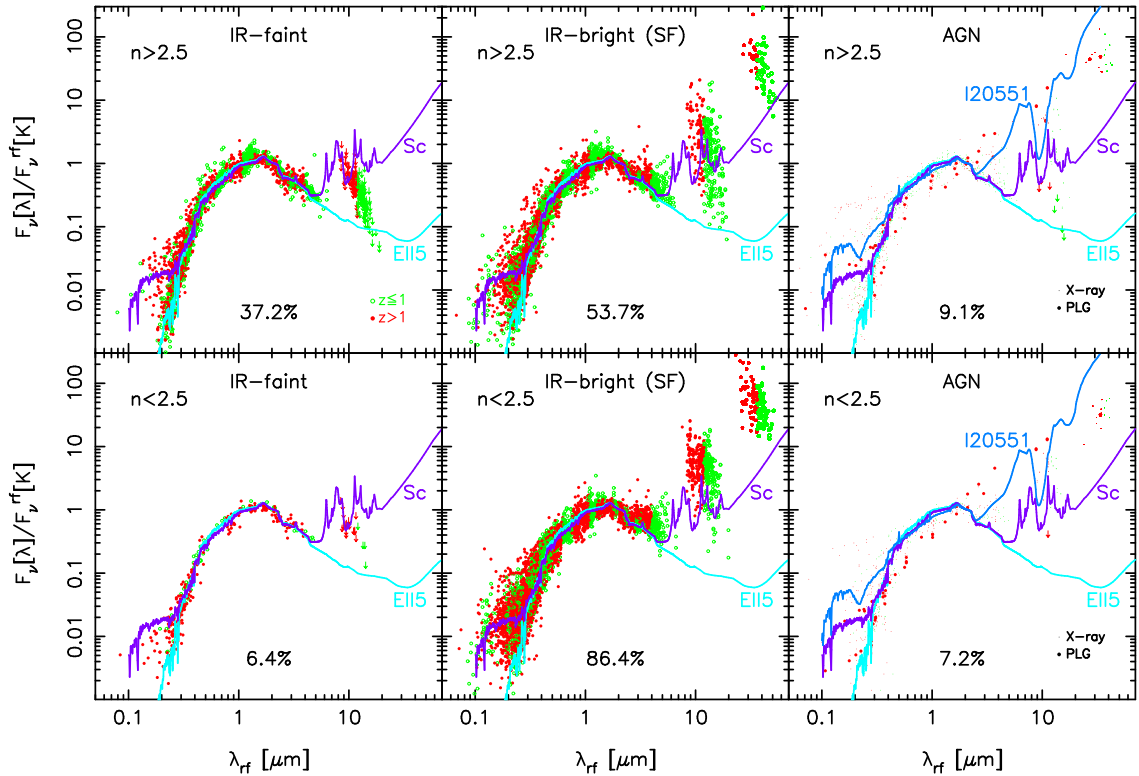}
\figcaption{\label{seds} Spectral energy distributions (de-redshifted 
and normalized to the rest-frame $K$-band flux) of all the galaxies in
the sample of massive galaxies
($\mathcal{M}$$>$10$^{11}$~$\mathcal{M}_\sun$) in the EGS
\citep{2007MNRAS.382..109T}. The upper and lower panels show the SEDs
for the spheroid-like and disk-like galaxies, respectively. For each
morphological type, the sub-sample has been divided in three groups:
galaxies without MIPS detection (IR-faint, left panels, with arrows
showing an upper limit of the MIPS 24\mic\, emission corresponding to
15\micJy, the minimum flux observed in the sample$^5$), galaxies with
a MIPS counterpart most probably linked to on-going star formation
(IR-bright, middle panels), and galaxies with nuclear activity (AGN,
right panels, see text for details). In each panel, open green circles
show sources at z$<$1, and filled red circles depict galaxies at
1$<$z$<$2. We also show typical templates (from
\citealt{2007ApJ...663...81P}) of an elliptical galaxy (Ell5), a
late-type spiral galaxy (Sc), and a galaxy with an obscured AGN
(I20551, just for the AGN panels on the right). All panels show the
fraction of sources in each type for the spheroid-like and disk-like
samples.}
\end{center}
\end{figure*}

Figure~\ref{seds} shows the SEDs of all the massive galaxies in our
EGS sample divided into morphological and activity types. The two
panels on the right show the SEDs for galaxies with an X-ray detection
(\citealt{2006ApJ...642..126B}, \citealt{2007ApJ...660L..11N}; see
also
\citealt{2007MNRAS.381..962C}) or classified as IRAC
power-law galaxies (PLGs, \citealt{2007ApJ...660..167D}), i.e.,
sources which most probably harbor an X-ray and/or IR emitting
AGN. There are 68 X-ray emitters, 60 of them with MIPS detection, and
5 PLGs in total, all of them with X-ray emission and 4 with MIPS
emission.


The SED distribution of the spheroids present a lower scatter in the
UV/optical, being very similar to a spectral template of an
elliptical. In contrast, there is a very populated tail of disk-like
galaxies with UV/optical fluxes brighter than a template for a typical
Sc galaxy, most probably linked to a recent starburst. The MIR
emission is consistent with the PAH spectrum of a late-type spiral
galaxy (see also Figure~\ref{postage_stamps}), but can be as high as
6-10 times the flux of the Sc template from
\citet{2007ApJ...663...81P}. Spheroid-like galaxies with a
MIPS detection present a lower 24\mic\, median flux (82\micJy, with
the quartiles being 42\micJy\, and 188\micJy, and the average
160\micJy) than MIPS disky sources (190$^{301}_{107}$\micJy, and the
average 250\micJy). Most of the sources identified as AGN present
relatively bright fluxes at rest-frame wavelengths between 2 and
10\mic, revealing the presence of very hot dust heated by the central
supermassive black hole and emitting in the NIR and MIR. In several
cases, this NIR/MIR emission hides the 1.6\mic\, bump, typically seen
in galaxies whose spectrum is dominated by stars rather than dust.


\subsection{Specific Star Formation Rates}


\begin{figure}
\begin{center}
\includegraphics[angle=-90,width=8.5cm]{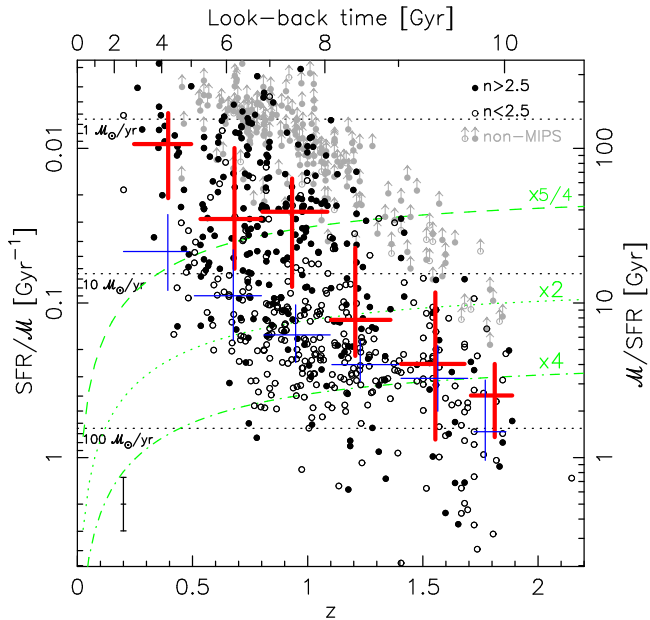}
\figcaption{\label{sfrm} Specific SFRs as a function of redshift and morphology 
(for galaxies not identified as bright AGN). Galaxies detected at
24\mic\, are plotted with open (disks) and filled (spheroids) black
circles, while gray symbols show upper limits for sources not detected
by MIPS. Red and blue crosses represent the median and quartiles for
the distribution of specific SFRs in the different redshift ranges
used in \citet[][red widest lines referring to spheroids and blue
narrowest to disky galaxies]{2007MNRAS.382..109T}. Green curves show
the expected positions of galaxies which would multiply their stellar
mass by 5/4, 2, and 4 between their redshift and z$=$0 if they
maintained a constant SFR. Horizontal dashed lines show constant SFR
values for the median stellar mass of our sample
(1.6$\times$10$^{11}$~$\mathcal{M}_\sun$).}
\end{center}
\end{figure}

Figure~\ref{sfrm} shows the specific SFRs of massive galaxies as a
function of redshift and morphology.  Table~\ref{table1} gives the
median and quartiles for different redshift ranges and morphological
types (obtained from S\'ersic index and visual classifications). The
median specific SFRs increase by less than 0.1dex and 0.02dex for
spheroids and disks, respectively, when considering MIPS 24\mic\,
detected above the 5$\sigma$ level. These increments do not affect the
following results significantly.

When segregating the sample based on the S\'ersic indices, we find
that the specific SFRs of spheroid galaxies evolve as
(1+z)$^{5.5\pm0.6}$ from z$=$0 to z$=$2, while the evolution for
disk-like galaxies goes as (1+z)$^{3.6\pm0.3}$. If we consider the
results based on the visual classification, the evolution is more
pronounced for spheroids and almost identical for disks:
(1+z)$^{6.4\pm0.8}$ evolution for the former, and (1+z)$^{3.4\pm0.2}$
for the latter.

The specific SFRs used in Figure~\ref{sfrm} have been estimated by
adding the unobscured SFRs obtained from UV data (at rest-frame
280~nm) and the obscured SFRs from IR data (using the total IR
luminosity) as explained in \citet{2008ApJ...675..234P}. The ratio
between these two quantities allow the estimation of the global
obscuration of the recent star formation in each galaxy (only for
those detected by MIPS). On average, we find that extinctions for MIPS
detected galaxies increase with redshift, ranging from
$<$$A(V)$$>$$=$1.0$\pm$0.5~mag at z$<$0.5 to
$<$$A(V)$$>$$=$1.5$\pm$0.6~mag at z$\sim$1 and
$<$$A(V)$$>$$=$2.0$\pm$0.7~mag at z$\sim$2. These values are
consistent with typical attenuations found for IR-bright galaxies by,
e.g., \citet{2000ApJ...537L..85R}, and the evolution in the extinction
properties of the UV SFR density found by
\citet{2007A&A...472..403T}. According to \citet{2007ApJ...660L..73S}, 
even larger extinctions (up to a factor of $\sim$100) are needed to
match SFRs obtained from IR or radio data and SFRs obtained from [OII]
spectroscopic observations of IR-bright sources in the EGS.


Figure~\ref{sfrm} shows that below z$=$1.1 spheroid-like galaxies
present very low specific SFRs. On average, they would increase their
stellar mass by less than 25\% at 0$<$z$<$1 if they maintained a
constant SFR. The global mass increase (in the form of newly-formed
stars) for all spheroid-like galaxies is less than 10\% if we take
into account the 38\% of z$<$1 spheroids which are not detected by
MIPS, and even lower ($\sim$5\%) if we only consider the visually
identified spheroids. In contrast, disk-like galaxies could typically
double their mass from z$=$1 to z$=$0 due to newly formed stars if
they maintained a constant SFR, with little change due to the very few
galaxies (less than 5\%) for which we only have SFR upper limits.

In practice, IR-bright intense star-forming bursts are not expected to
last long \citep{1994ApJ...431L...9M}, so a galaxy most probably will
not maintain a high SFR level for several Gyr. The higher SFR values
are expected to be maintained for shorter periods, since gas
exhaustion and supernova winds (and even AGN activity) will help to
suppress star formation. \citet{1996ApJ...464..641M} simulations of
mergers predict the triggering of a very intense and short starburst
event (probably detectable in the MIR by MIPS) lasting a few tens of
Myr (and occurring in late stages of the merger, when the galaxies are
actually joining) for encounters of galaxies with an already formed
bulge. Encounters of disky galaxies would trigger less intense bursts
lasting longer (100-200~Myr) and occurring earlier in the merger
process. For the observed specific SFRs in our sample, those short and
intense starbursts would add up less than 0.01\% (for each merging
event) to the total stellar mass of a typical spheroid-like galaxy at
z$<$1. This very small fraction of young stars would be hidden by the
predominant old stellar population and be undetectable in local
ellipticals. For the disky galaxies at z$<$1, which present specific
SFRs as high as 0.2~Gyr$^{-1}$, the burst strength (ratio of the newly
formed stars to the global stellar mass) could be as high as a few
percent (for each merging event), typical for star-forming galaxies at
low-redshifts \citep{2003MNRAS.341...33K,2003MNRAS.338..525P}.

At z$\gtrsim$1, the specific SFRs of massive galaxies are higher than
0.1~Gyr$^{-1}$, both the active spheroid-like (note that there are
40\% of spheroids which are not detected by MIPS) and disky systems
are forming stars at approximately the same rate, and the number of
quiescent galaxies (those not detected by MIPS) is less than
$\sim$50\% for both types. It is interesting to notice that most
galaxies (disks and spheroids) have significant amounts of dust, since
they are detected at 24\mic. If some of these galaxies are the
progenitors of nearby ellipticals, that dust should have disappeared
somehow or it is now very cold and may only be detected at very long
wavelengths ($\lambda$$>$100-200\mic) and low fluxes.

For typical burst durations, and even for star-forming events with a
constant SFR and lasting up to 1~Gyr, the maximum increase in stellar
mass would be $\sim$15\% at 1.1$<$z$<$1.4, $\sim$25\% at
1.4$<$z$<$1.7, and $\sim$50\% at 1.7$<$z$<$2.0, for both spheroids and
disks. This means that a significant fraction (more than 50\%) of the
stellar mass of z$>$1 massive galaxies was assembled at z$>$2
(\citealt{2008ApJ...675..234P}; see also
\citealt{2005ApJ...621L..89B}, \citealt{2005ApJ...633L...9F}, 
\citealt{2006ApJ...640...92P}). Moreover, we find that about 40\% of spheroids
at z$\sim$1.8 are almost ``dead'' (they present low SFR levels based
on IR and UV data) and evolving passively, or may be experiencing a
quiescent period. Note that most spheroid-like galaxies at z$>$1 would
be qualified as passive based on optical colors alone, but the MIPS
data reveals that $\sim$50\% of them are experiencing dusty starbursts
and 10\% more harbor (also) obscured AGN.

We can estimate how much stellar mass galaxies typically assemble
through star formation from z$=$2 to the present (i.e., the star
formation efficiency to increase the mass of a galaxy in the last
10~Gyr) if a galaxy follows the specific SFR evolution depicted in
Figure~\ref{sfrm}. We assume that the SFRs remain constant within each
redshift interval; since the starbursts probably last 50-200~Myr, as
discussed earlier, the following figures would be an upper
limit. Adding all the mass formed from z$=$2 to z$\sim$0, we estimate
that a disk-like galaxy could increase its stellar mass by up to a
factor of 3.2$\pm$0.5 in the last 10~Gyr: 1.4 times increase at
1.7$<$z$<$2.0 and an almost constant 10-20\% increase in each of our 5
redshift intervals at z$<$1.7. A spheroid-like galaxy could increase
its stellar mass by up to a factor of 1.8$\pm$0.3: 1.2 times increase
at 1.7$<$z$<$2.0, 10-20\% in each of our two intervals at
1.1$<$z$<$1.7 and less than 5\% in each of the three intervals at
z$<$1.1. These figures are almost unchanged ($<$5\% increases) when
considering only the MIPS 24\mic\, 5$\sigma$ detections. For visually
identified disks and spheroids, the stellar mass increases by up to a
factor of 2.7$\pm$0.4 and 1.8$\pm$0.3, respectively.

Ideally, one would like to disentangle what is the relative
contribution to the size growth of a galaxy of newly-formed stars and
system heating through merger/interactions. However, both processes
are probably linked, since new star formation events are likely
associated to the interactions that could inject energy to the
systems. Consequently, a definitive answer to the problem of how
galaxies grow require the help of elaborate modelling. It is worth
saying, nonetheless, that both the observations at low and high-z show
that galaxies have larger effective radii when observed in bluer
bands. This means that younger stars are preferentially located at
larger galactocentric distances than older populations. In this paper,
we have quantified how much the stellar mass grows through star
formation events only. Once we reach a clear picture of how the
galaxies can increase in size through mergers, our results will
constrain the amount of stellar mass due to dry accretion that is
necessary to migrate the high-z galaxies to the local size--mass
relations.

\section{Summary and conclusions}

We have analyzed the stellar mass growth in the form of newly-born
stars in a sample of 831 $K$-band selected massive galaxies
($\mathcal{M}$$>$10$^{11}$~$\mathcal{M}_\sun$) as a function of
structural parameters (size and concentration). These galaxies lie in
the redshift range between z$=$0.2 and z$\sim$2. Our analysis is based
on the measurement of the specific SFR for each galaxy based on their
UV and IR emission, taking advantage of the deep
\spitzer\, data obtained by the FIDEL \spitzer/MIPS Legacy Project in the 
Extended Groth Strip.

Our main results follow:

\begin{itemize}
\item[$-$]Most (more than 85\% at any redshift) disk-like galaxies 
(identified by small S\'ersic indices, $n$$<$2.5) are detected by MIPS
at 24\mic\, down to $F(24)$$=$15\micJy\, with a median flux
$F(24)$$=$190\micJy.
\item[$-$]A significant fraction (more than 55\% at any redshift) of 
spheroid-like galaxies is detected at 24\mic\, down to
$F(24)$$=$14\micJy\, with a median flux $F(24)$$=$82\micJy.
\item[$-$]The MIPS detection fraction for spheroid-like galaxies is higher 
(70--90\%) for larger ($r_e$$\gtrsim$5~kpc) galaxies, especially at
z$>$1, where the detection fraction has a minimum around 30--40\% for
galaxies with $r_e$$\sim$1~kpc. No clear trend is found for disky
galaxies of different sizes.  
\item[$-$]There is basically no difference between the loci occupied by 
MIPS detected and undetected galaxies in the stellar mass--size
plane. However, for spheroid-like objects at a given stellar mass,
MIPS non-detections are smaller than IR-bright sources by a factor of
$\sim$1.2.
\item[$-$]Most of the galaxies in our sample present spectral energy 
distributions which are consistent with an elliptical template in the
UV/optical/NIR spectral range. Some galaxies morphologically
classified as spheroids have UV emission tails which are typical of
star-forming systems, most commonly at z$>$1. 
\item[$-$]A $\sim$10\% fraction of the massive galaxies in our sample present X-ray or 
power-law-like mid-IR emission which must be linked to the presence of
a bright (unobscured or obscured) AGN.
\item[$-$]Based on the measured specific SFRs, we estimate that spheroid-like 
galaxies have doubled (at the most, depending on the burst durations)
their stellar mass due to newly-born stars alone from z$\sim$2 to
z$=$0.2. Most of these mass increase (60\%) occur at z$\gtrsim$1,
where specific SFRs are as high as 0.4~Gyr$^{-1}$.
\item[$-$]Disk-like galaxies have tripled (at the most) their stellar mass by 
newly-formed stars at z$<$2, with a more steady growth rate as a
function of redshift.
\end{itemize}

\acknowledgments 

We thank an anonymous referee for her/his very constructive
comments. We acknowledge support from the Spanish Programa Nacional de
Astronom\'{\i}a y Astrof\'{\i}sica under grants AYA 2006--02358 and
AYA 2006--15698--C02--02. This work is based in part on observations
made with the {\it Spitzer} Space Telescope, which is operated by the
Jet Propulsion Laboratory, Caltech under NASA contract 1407. PGP-G and
IT acknowledge support from the Ram\'on y Cajal Program financed by
the Spanish Government and the European Union.

\bibliographystyle{apj}
\bibliography{referencias}

\end{document}

%% file: tab1.tex
\begin{deluxetable*}{lrrccccrrccc}
\tabletypesize{\tiny}
\tablewidth{500pt}
\tablecaption{\label{table1}MIPS detection fraction and specific SFR statistics as a function of morphology.}
\tablehead{ & \multicolumn{5}{c}{S\'ersic index classification} & & \multicolumn{5}{c}{Visual classification}\\
\cline{2-6} \cline{8-12} \\
\colhead{Redshifts} & \multicolumn{2}{c}{Sources (MIPS detected)} & & \multicolumn{2}{c}{log(SFR/$\mathcal{M}$$^\dag$)} & & \multicolumn{2}{c}{Sources (MIPS detected)} && \multicolumn{2}{c}{log(SFR/$\mathcal{M}$$^\dag$)}\\
\cline{2-3} \cline{5-6} \cline{8-9} \cline{11-12}\\
 & \colhead{n$\leq$2.5} & \colhead{n$>$2.5} & & \colhead{n$\leq$2.5} & \colhead{n$>$2.5} & & \colhead{S/Irr/mergers} & \colhead{E/S0} & & \colhead{S/Irr/mergers} & \colhead{E/S0}
}
\startdata
(0.2,0.5]  &   8 (100.0\%) &  31 (80.6\%) &  & -1.33$^{-1.57}_{-1.08}$ & -2.03$^{-2.23}_{-1.68}$  &  &  14 (100.0\%) &  25 (76.0\%) &  & -1.37$^{-1.68}_{-1.12}$ & -2.12$^{-2.36}_{-2.01}$ \\
(0.5,0.8]  &  55  (94.5\%) & 144 (60.4\%) &  & -1.05$^{-1.34}_{-0.76}$ & -1.54$^{-2.00}_{-1.22}$  &  &  88  (94.3\%) & 111 (50.5\%) &  & -1.12$^{-1.43}_{-0.79}$ & -1.82$^{-2.07}_{-1.46}$ \\
(0.8,1.1]  & 120  (95.8\%) & 173 (57.2\%) &  & -0.79$^{-0.99}_{-0.62}$ & -1.59$^{-1.81}_{-1.10}$  &  & 156  (95.5\%) & 137 (48.2\%) &  & -0.86$^{-1.06}_{-0.63}$ & -1.63$^{-1.86}_{-1.44}$ \\
(1.1,1.4]  &  83  (94.0\%) &  83 (66.3\%) &  & -0.60$^{-0.76}_{-0.49}$ & -0.89$^{-1.36}_{-0.66}$  &  & 104  (98.1\%) &  62 (50.0\%) &  & -0.63$^{-0.83}_{-0.49}$ & -1.26$^{-1.48}_{-0.90}$ \\
(1.4,1.7]  &  56  (87.5\%) &  40 (57.5\%) &  & -0.51$^{-0.72}_{-0.30}$ & -0.61$^{-1.07}_{-0.12}$  &  &  60  (86.7\%) &  36 (52.9\%) &  & -0.51$^{-0.77}_{-0.23}$ & -0.61$^{-0.98}_{-0.32}$ \\
(1.7,2.0]  &  21  (85.7\%) &  12 (58.3\%) &  & -0.17$^{-0.50}_{+0.02}$ & -0.40$^{-0.61}_{-0.13}$  &  &  24  (83.3\%) &   9 (50.0\%) &  & -0.36$^{-0.61}_{+0.01}$ & -0.17$^{-0.33}_{-0.06}$ \\
\enddata
\tablecomments{$^\dag$ In units of Gyr$^{-1}$.}
\end{deluxetable*}